\documentclass[twoside]{article}
\input ibvs3.sty

\usepackage{epsfig}
\usepackage{epstopdf}
\usepackage{graphicx}
\usepackage{rotating}
\usepackage{footnote}

\newcommand{\etal}{$et~al.~$}

\begin{document}
\IBVSheadDOI{63}{6245}{20 July 2018}

\IBVStitle{The Period Evolution of V473\,Tau}

\IBVSauth{OZUYAR, D.$^1$; STEVENS, I. R.$^2$}

\IBVSinst{Ankara University, Faculty of Science, Dept. of Astronomy and Space Sciences, 06100, Tandogan, Ankara / Turkey, e-mail: dozuyar@ankara.edu.tr}

\IBVSinst{The University of Birmingham, School of Physics and Astronomy, Birmingham, B15 2TT, UK}

\SIMBADobjAlias{V473\,Tau}{HD 30466}

\IBVSabs{In this paper, the period evolution of the rotating chemically peculiar star V473\,Tau is investigated. Even though the star has been observed for more than fifty years, for the first time four consecutive years of space-based data covering between 2007 and 2010 are presented. The data are from the {\sl STEREO} satellite, and are combined with the archival results. The analysis shows that the rotation period of V473\,Tau is $1.406829(10)$ days, and has slightly decreased with the variation rate of 0.11(3)~s~yr$^{-1}$ over time. Also, the acceleration timescale of the star is found to be shorter than its main sequence lifetime. This indicates that the process of decrease in period might be reversible. On this basis, it can be suggested that V473\,Tau has a possible magnetic acceleration and a differential rotation, which cause a variation in the movement of inertia, and hence the observed period change. Additionally, the evolution path of V473\,Tau on the H-R diagram is evaluated. Accordingly, the position of the star on the diagram suggests that its magnetic properties develop before it reaches the main sequence or in the beginning of its main sequence lifetime.}

\begintext

\section{Introduction}
\label{intro}

Chemically peculiar (CP) variables are spread between late-B and early-F spectral types, and thus contain various stars with effective temperatures greater than 6,500 K (Hubrig \etal 2005). These variables are comprised mostly of Ap and Bp stars, which differ from other types having the same temperature by their abnormal chemical compositions and slow rotations. The reason for the peculiarity is an under-abundance of solar-like elements, as well as an overabundance of both metal and rare-earth elements across their surfaces (Mikulasek \etal 2009). Magnetic fields, radiative acceleration, and atomic diffusion determine the surface distribution of elements (Kochukhov 2011), and lead them to be present in the form of spots and rings on the surface. Along with rotation, these non-uniformly distributed regions cause periodic variations in magnetic fields, line profile, and energy distribution, as well as in photometric brightness (oblique-rotator model). The periods of these variations are generally between a day and a week. Depending on the slow rotation, surface spot regions can remain stable for decades. Such a situation enables remarkably accurate calculations of surface distribution, rotation period, and rotational breaking mechanisms. However, only very few of the CP stars discovered in our galaxy and others exhibit periodic variations, and less than one-tenth of these have been observed for scientific investigation. In order to study this type of stars, accurate observations are needed (accuracy $>$ 0.005 mag; Mikulasek \etal 2009). The high-precision instruments of the {\sl STEREO} satellite are a quite suitable, space-based source, since seasonal and four-year {\sl STEREO} observations provide a precision of $2.0 \times10^{-4}$ and $7.0 \times 10^{-5}$ mmag, respectively.

\section{Literature Review}
\label{review}

The photometric variability of V473\,Tau (A0Si, V = 7.26 mag) was first detected by Burke \etal (1970). They calculated the period of this variation as around 1.39(2) days, but this period value produced a light curve (LC) with a scattered maximum. Hence, Rakosch and Fiedler (1978) noted that their observations were more adaptable with a double period. Subsequently, Maitzen (1977) derived a rotation period of 2.7795(1) days, which was indeed twice that of previous values. As a result of the double period, two minima and maxima having different levels were formed in the LC; this situation was explained in terms of the different chemical regions on the surface. Most importantly, this was a significant case since a double wave structure was not a common condition among Si stars. In a recent study, Jerzykiewicz (2009) investigated rotation periods and found a value of 1.4068541(29) days in U, B, and V bands. However, he could not completely determine the origin of the variabilities as he was unable to conclude whether the star was an oblique rotator or a g-mode pulsator.

\section{Analysis of the STEREO Data}

The data were provided from the HI-1A instrument on-board the {\sl STEREO-A} satellite. The HI-1A is capable of observing background stars with the magnitude of $12^m$ or brighter for a maximum of 20 days and a useful stellar photometer which covers the region around the ecliptic (20\% of the sky) with the field of view of $20^{\circ} \times 20^{\circ}$. The nominal exposure time of the camera is 40 seconds, and putting 30 exposures together on board, a 40-minute integrated cadence has been obtained to transmit for each HI-1 image (Eyles \etal 2009). Therefore, the Nyquist frequency of the data is around 18~c~d$^{-1}$. LCs mostly affected by solar activities were cleaned with a $3^{rd}$ order polynomial fit. Observation points greater than $3 \sigma$ were clipped with a pipeline written in the Interactive Data Language (IDL) (For a more detailed description of the data preparation, refer to Sangaralingam and Stevens (2011) and Whittaker \etal (2012)). The LC of V473\,Tau presented a sinusoidal characteristic due to spot modulation on the stellar surface. Therefore, all analyses were performed using the Lomb-Scargle (LS) algorithm since it is more sensitive to such variations. To determine a model of the sinusoidal LCs, the Levenberg-Marquardt Optimization method was applied, and the best fit was obtained after 5000 iterations. After the derivation of the model LC, random Gaussian noise with the mean of zero and the sigma value, which was determined from the cleaned curve, was produced and added to the model. This process was repeated 500 times. The most accurate frequencies and their uncertainties were assessed using the Monte-Carlo simulation algorithm. The results were compared to those derived from the Phase Dispersion Minimization (PDM) method and Period04. To perform O-C calculations and to investigate period variabilities over years, the best extremum times  were obtained from the seasonal LCs, and were put together with data from the literature.

\begin{table}[!htp]
\centerline{{\bf Table\,1.} Frequency analysis results of V473\,Tau.}
\vskip 3mm
\begin{center}
\begin{tabular}{l l l l c}
\hline\hline
V473\,Tau&LS&Period04&PDM&Amp.\\
&(c~d$^{-1}$)&(c~d$^{-1}$)&(c~d$^{-1}$)&(mmag)	\\
\hline																	
2007	&	0.7104(8)	&	0.7101(8)	&	0.7120(16)	&	8.63(25)	\\
2008	&	0.7101(6)	&	0.7101(6)	&	0.7118(15)	&	10.93(24)	\\
2009	&	0.7116(7)	&	0.7116(8)	&	0.7157(15)	&	8.62(24)	\\
2010	&	0.7128(7)	&	0.7128(8)	&	0.7123(20)	&	9.20(25)      \\
Comb.	&	0.710818(5)	&	0.710818(7)	&	0.711164(5)	&	9.33(13)	\\
\hline \hline
\end{tabular}
\end{center}
\end{table}

\section{Results}
\label{results}

In this research, we obtained four consecutive years of data between 2007 and 2010. As reported by other researchers, all the LCs had explicit periodicity. Individual  LS, PDM and Period04 analyses of annual curves showed a frequency at around 0.71 c~d$^{-1}$ (Table~1), but this result was slightly longer than the literature periods. Furthermore, we detected the existence of another strong peak at approximately 1.40 c~d$^{-1}$ (0.71 days) on the LS periodogram (Figure~1).

\vskip 0.5cm
\begin{table}[!hp]
\centerline{{\bf Table\,2.} Available period values and extremum times for V473 Tau.}
\vskip 3mm
\begin{center}
\begin{tabular}{c l l l l l}
\hline \hline          
\textbf{Time} \textbf{(year)}	&	\textbf{Period} \textbf{(day)}	&\textbf{Frequency} \textbf{(c~d$^{-1}$)} 	&\textbf{Ref.}&\textbf{Extremum Times} \textbf{(HJD)}&\textbf{Ref.}\\	

\hline
1963-1993 & 1.4068541(29) & 0.710806(1)& 1 &			2438451.1380(100) & 1 \\	
1967-1968 & 1.39(2) & 0.72(1)&2 &			2438451.1540(220) &1 \\			
1963-1964 & 1.39 &0.72 &3  &			2438750.7800(190) & 1\\			
1974 & 1.38975(5) & 0.71955(3)&4 &					2439860.8060(230) &1 \\			
1990-1993 & 1.4066952 & 0.710886&5  &				2448480.6010(190) & 1\\		
1990-1993 & 1.407020(39) &0.7107(2) &6  &			2439870.6300(500) &  2  \\
2007 & 1.4069(6) &0.7108(3)&7  &				2438466.7297 & 3 \\	
2007-2010 & 1.406829(10) &  0.710818(5)&8  &		2438466.3665(1300) & 4 \\				
		&&	&&		2454241.5599(125) & 8\\			
		&&	&&		2454583.4049(129) & 8\\			
		&&	&&		2454922.4465(133) & 8\\			
		&&	&&		2455274.1565(135) & 8\\	

\hline  \hline    
\end{tabular}
\raggedright \\
1: Jerzykiewicz (2009), 2: Burke \etal (1970), 3: Rakosch \& Fiedler (1978) \\
4: Maitzen (1977) (P/2), 5: Dubath \etal (2011), 6: Rimoldini \etal (2012)  \\
7: Wraight \etal (2012), 8: This study \\
\end{center}
\end{table}

Combining the four-year data, the precise rotation period of the star was determined with the help of the PDM and LS methods. Since the LS technique gave a better period, the main LC was plotted based on this value. Accordingly, the folded LC was clearly formed by a maximum and a broad minimum (Figure~2, upper left). The maximum was quite strong and had a flat top, indicating a cooler chemical structure on the surface. Moreover, there was a barely detectable bump in the middle of the minimum. From the Figure~2, it was clear that the light curve did not have a purely sinusoidal shape. As a result of this, it produced a Fourier spectrum comprised of an $nf$ ($n=1, 2, 3, \ldots$) harmonic series with decreasing amplitudes with increasing $n$. Therefore, the peak at 1.40 c~d$^{-1}$ on the LS periodogram was the first harmonic of the main frequency.

Also, we produced a folded LC using the double {\sl STEREO} period since Maitzen (1977) noted that his observations were compatible with the period value of 2.7795 days. As shown in Figure~2 (upper right), we derived a relatively clean LC with two minima and maxima. Even though the minimum at $\phi \approx 0.3$ was slightly more scattered than the other one, the consecutive structures appeared similar to each other. Therefore, we assumed that the period value of 1.41 days was the full rotation period.

To investigate a possible period variation, we collected all literature values given in Table~2, and present them in Figure~2 (bottom left) using black diamond symbols. Since some of them were the results of multi-observations, we used the combined {\sl STEREO} period instead of seasonal periods (a red diamond symbol). As seen in the figure, we found two different period paths ($\approx$1.390 and $\approx$1.408 days) since the quality and number of observation data differed from one study to another. Therefore, it was not possible to calculate any period variation using these values. However, when only the values given in Jerzykiewicz (2009) (10-year observations), Wraight \etal (2012), and this study ({\sl STEREO} observations) were considered based on their reliabilities, the change in period suggested a possible period increase with a rate of 0.03 s~y$^{-1}$  in the star over 45 years. 	

In order to confirm such a variation, we analysed variabilities in the O-C diagram. For the calculation, the maximum times of the individual LCs were derived, and these values were combined with the epochs from the literature, given in Table~2. The epochs provided by Rakosch and Fiedler (1978), and Maitzen (1977) were converted from JD to HJD. Based on  Figure~2 (right bottom), we found out that the O-C diagram of the star exhibited a period decrease with the variation rate of around $-1.27(30) \times 10^{-6}$ d~y$^{-1}$  or $-0.11(3)$ s~y$^{-1}$ (blue straight line). With the help of the LS period and using the best {\sl STEREO} maximum time, we determined the light elements as:   

\begin{equation}
\nonumber \\
HJD_{max} = 2454583.4049(129)+ 1.406829(10) E - 2.44(58) \times 10^{-9} E^2 ~ .
\end{equation}

Since this star was a single rotating variable, such a period decrease might most likely be explained by an acceleration in rotation after a magnetic braking, and might affect the dynamic structure of the star.  Using the physical parameters $T$ = 11,081(280) K, $M = 2.59(14)$ M$_{\odot}$, $log(L/$L$_{\odot}) = 1.64(15)$, and $R = 1.80(32)$ R$_{\odot}$, which was calculated from temperature and luminosity values provided, as given by Wraight \etal (2012), we roughly calculated the kinetic energy of the star and the rate at which energy increased as $E= 4.31(1.57) \times 10^{46}$ erg and $dE/dt = 2.46(1.07) \times 10^{33}$ erg~s$^{-1}$. We also found the corresponding angular momentum and its variation rate to be around $J = 1.67(61) \times 10^{51}$ erg\,s and $dJ/dt = 4.77(2.07) \times 10^{37}$ erg. According to period and angular momentum variations, the acceleration time-scale of the star was approximately $\tau_{AC} = 1.11(63) \times 10^{6}$ yr, which was slightly higher than the duration derived from the variation rate of the kinetic energy ($\Delta \tau = E/(dE/dt) = 5.55(3.15) \times 10^{5}$ yr). We also found the main sequence lifetime of the star as $\tau_{MS} = 9.26(1.25) \times 10^{8}$ yr from the equation of $\tau_\mathrm{MS} = 10^{10}~\mathrm{yr} \times (M/M_{\odot})^{(1-\alpha)}$, where $\alpha = 3.5$ for main sequence stars and $10^{10}$~yr is the approximate lifetime of the Sun in the main sequence (Ghosh 2007; Koupelis and Kuhn 2007; Hansen and Kawaler 1994).

In addition to these, such a period decrease might be a result of a change in stellar mass with a rate of around $dM/dt = -1.92(88) \times 10 ^{-12}$ M$_{\odot}$ yr$^{-1}$, or a consequent of a change in radius with a rate of around $dR = -8.10(2.42) \times 10^{-7}$ R$_{\odot}$ yr$^{-1}$. Finally, we found the rotational velocity of the star to be $V_{eq}= 65(12)$ km~s$^{-1}$ with the help of our combined LS period and radius value ($R = 1.80(32)$ R$_{\odot}$), estimated from the parameters given above.

\IBVSfig{20cm}{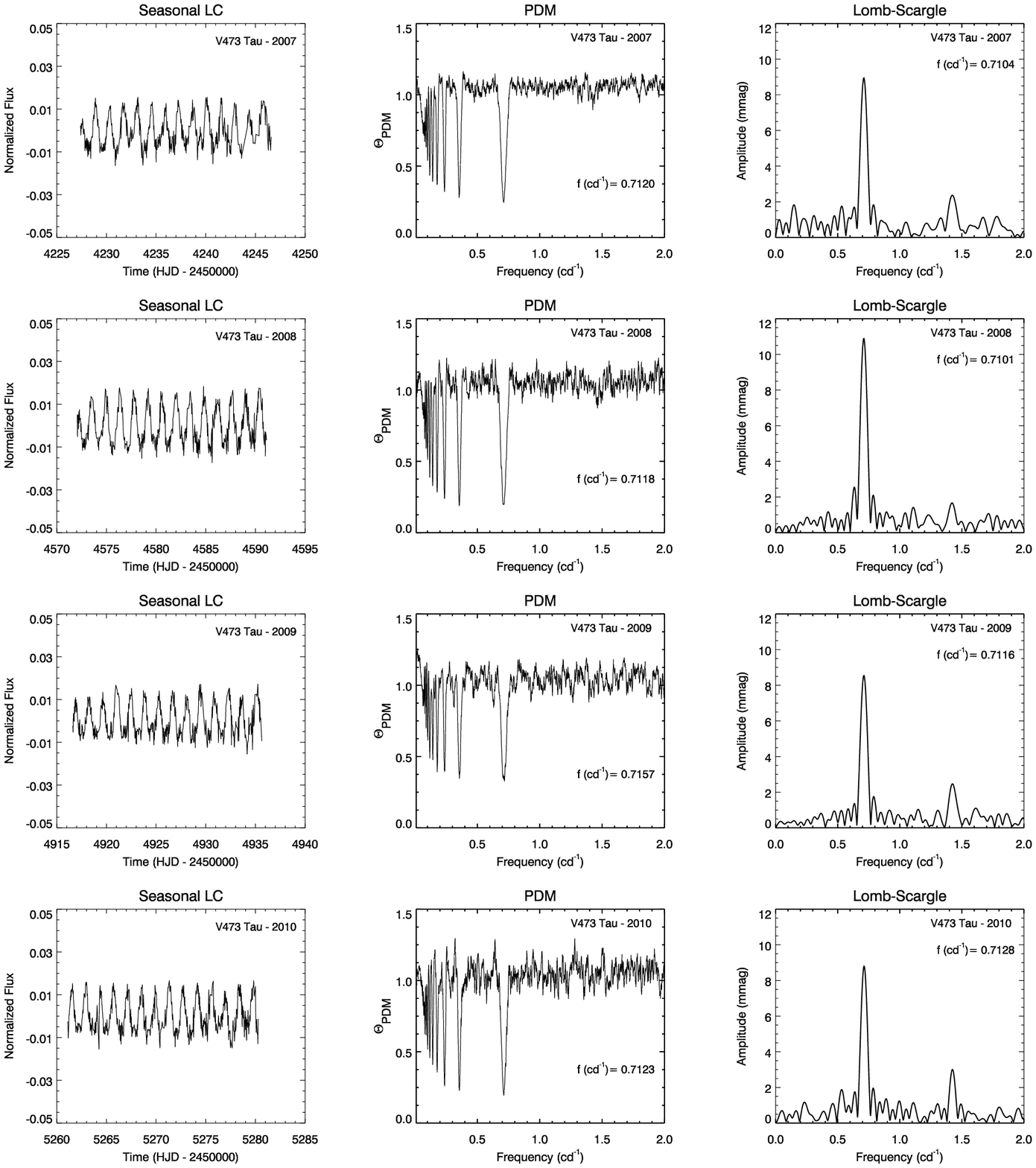}{Annual light curves and related frequency periodograms of V473\,Tau.}
\IBVSfigKey{f1.eps}{V473\,Tau}{light curves}

\IBVSfig{18cm}{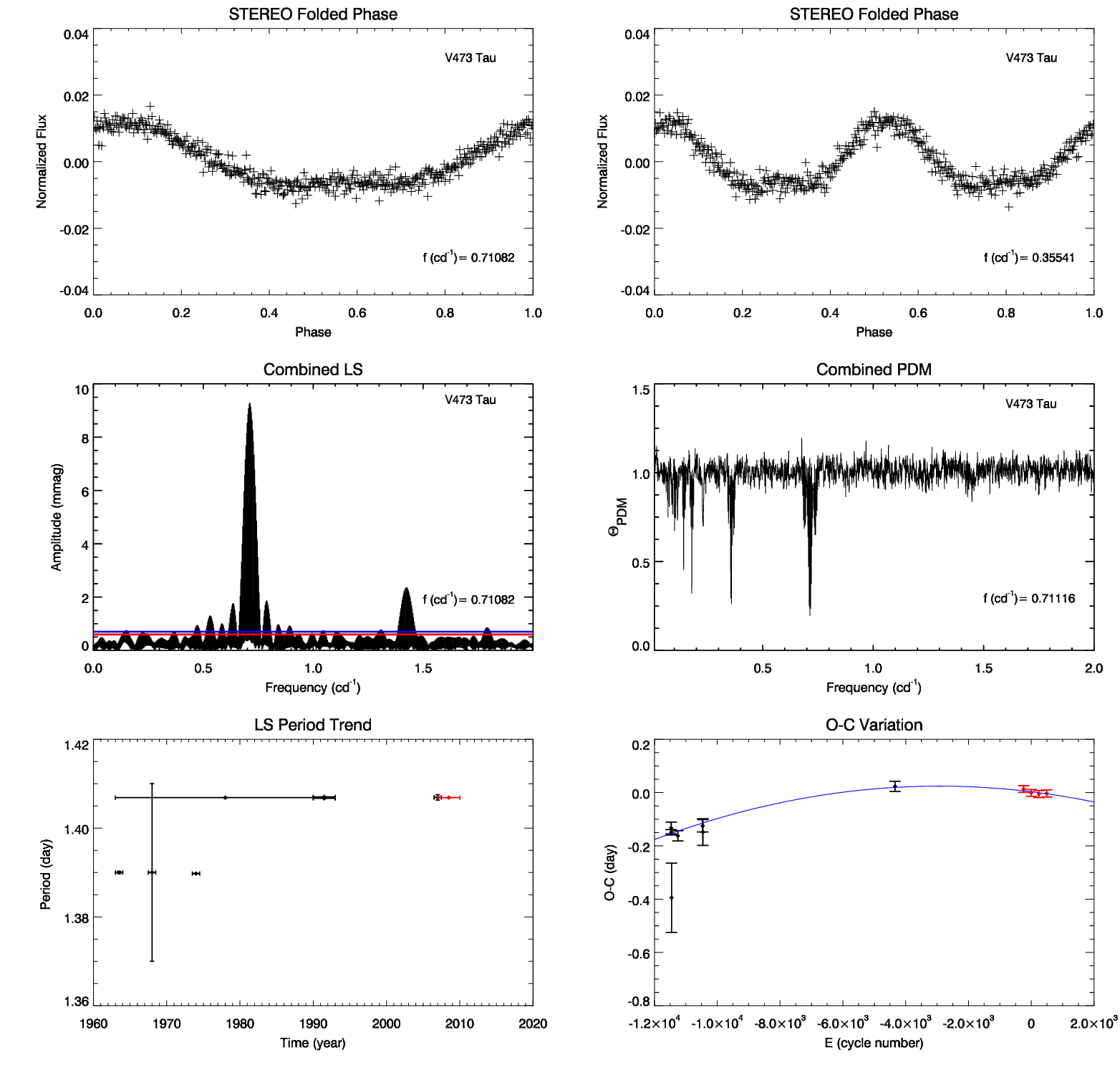}{Folded light curves produced by the {\sl STEREO} periods, frequency analyses of combined light curves as well as period and O-C variation graphics V473\,Tau.}
\IBVSfigKey{f2.eps}{V473\,Tau}{diagrams}

\vskip 1cm
\centerline{{\bf Table\,3.} The period, period variation rate, acceleration and main sequence lifetime}
\centerline{as well as physical parameters of V473\,Tau.}
\vskip 3mm
\begin{center}
\small
\begin{tabular}{c c c c c c} \hline\hline

$P$	&	$dP/dt$	&	$\dot{P}/P$&	$\tau_{ACC}$	&	$\tau_{MS}$\\
(day)&	(s~yr$^{-1}$)	&	(s$^{-1}$)	&	(yr)	&	(yr)	\\

\hline

1.406830(10)	&	-0.11(3)	&	-2.86(68)$\times 10^{-14}$	&	-1.11(63)$\times 10^{6}$	&	9.26(1.25)$\times 10^{8}$	\\

\hline\hline

$Log (L/$L$_{\odot})$&	$Log (T)$	&	$M$	&	$R$	&	$V_{eq}$	\\
   	& &	(M$_{\odot}$)	&	(R$_{\odot}$)	&	(km~s$^{-1}$)		\\\hline

1.64(15)	&	4.045(11)	&	2.59(14)	&	1.80(32)	&	65(12)	\\
\hline\hline
\end{tabular}
\end{center}

\section{Discussion}
\label{discussion}

V473\,Tau shows explicit period variation in the O-C diagram. Based on the diagram, it has been rotating 0.11(3) seconds faster per year. The variation rate in the period ($\dot{P}/P = 10^{-14}$ s$^{-1}$) is 10 times greater than that of the most massive mCP stars (Mikul{\'a}{\v s}ek \etal 2014). In addition, its acceleration time-scale is around $\tau_{MS} \sim 10^6$ yr, which is nearly three orders of magnitude ($\sim 0.8 \times 10^3$ yr) shorter than the main sequence lifetime of the star ($\tau_{MS} = 10^8$ yr). This, in turn, suggests that process of decrease in the period may be reversible. If so, the length of the cycle is roughly calculated as 92(11) yr (estimated by $T_{cyc} \sim P~\sqrt{2/\dot{P}}$, Mikul{\'a}{\v s}ek \etal (2010)). Considering the fact that period variation processes may be reversible due to shorter acceleration time-scale than that of the main sequence lifetime, the rigid rotation hypothesis should be discarded and the differential rotation model should alternatively be discussed as expressed by St{\c e}pie{\'n} (1998). In this model, the outer layers of stars differentially rotate with respect to denser interiors, and they are affected by global magnetic fields; an interaction between meridional circulations and magnetic fields takes place in a region within a star. 

\IBVSfig{8cm}{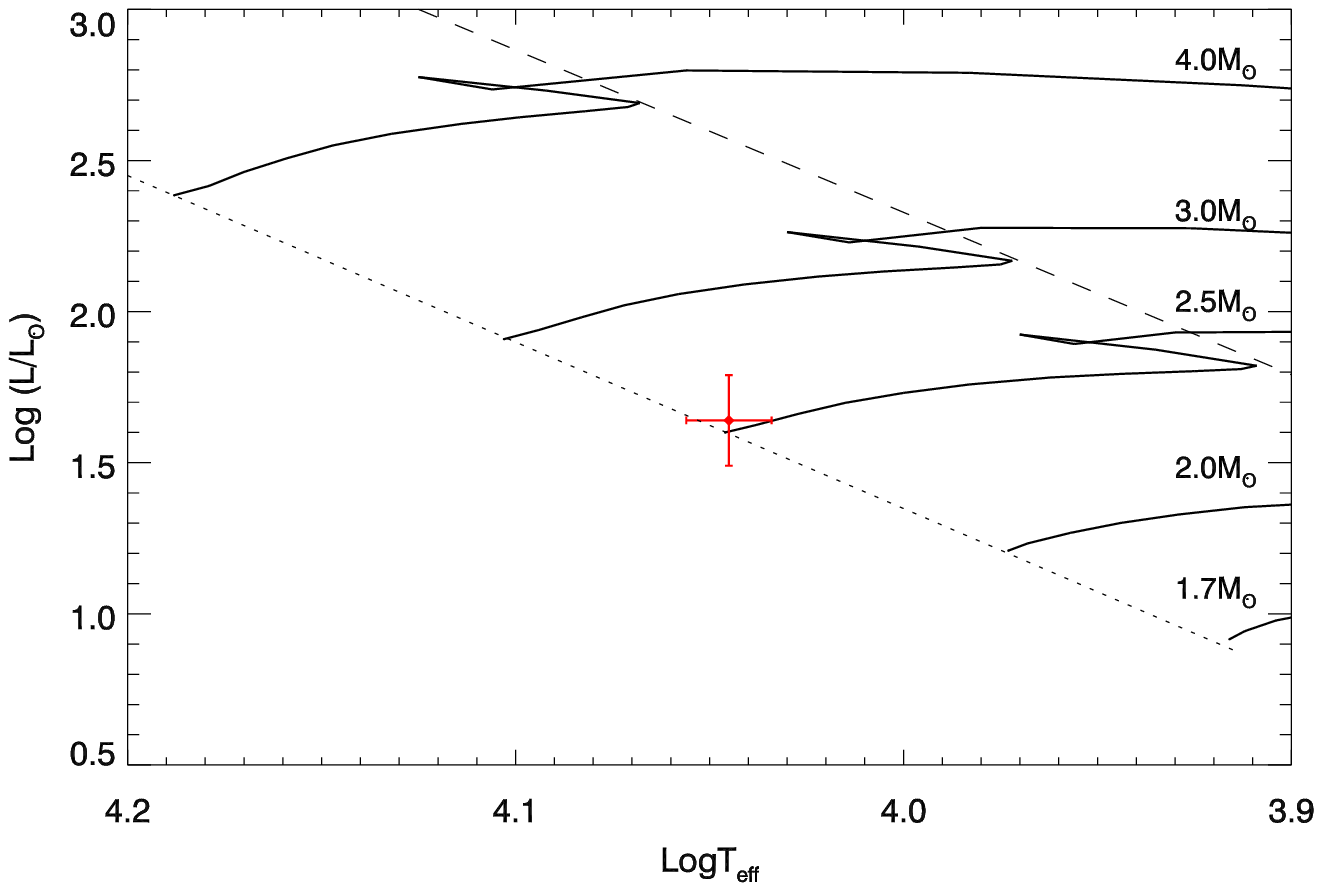}{Positions of V473\,Tau on the H-R diagram. Evolution
paths for intermediate mass stars (continuous lines), zero age main sequence (dotted line), and terminal
age main sequence lines (dashed line) are from Schaller \etal (1992).}
\IBVSfigKey{f3.eps}{V473\,Tau}{evolution graph}

This region is an interface between inner layers where circulation is dominant and the outer envelope is influenced by magnetism. As a result of differential rotation, a toroidal component of the internal magnetic field is produced, and it increases until the outer magnetically-confined envelope is forced to co-rotate with the interior. Hence, a cyclic increase and decrease in the moment of inertia occurs St{\c e}pie{\'n} (1998). This means that an unexpected alternating variability of rotation periods can be observed. In this case, rotation acceleration in V473\,Tau may be interpreted as a consequence of torsional oscillations produced by meridional circulations being in interaction with a magnetic field, and of rotational braking in outer layers caused by angular momentum loss via magnetically-confined stellar wind.

Additionally, the evolution track of the star on the H-R diagram is evaluated in this study (Figure~3). The temperature and luminosity values of the star are taken from Wraight \etal (2012). In Figure~3, evolution path for intermediate mass stars (continuous lines), zero age main sequence (dotted line), and terminal age main sequence lines (dashed line) are derived from Schaller \etal (1992). Based on the figure, the star is located close to the zero age main sequence, where its mass value is compatible with the theoretical evolution path. 

Oetken (1985), Hubrig and Mathys (1994) state that the magnetism of CP stars develops in the final stages of main sequence evolution. Also, Hubrig et al. (2000) indicate that magnetic fields show up only in stars that complete at least 30\% of their main sequence lifetimes. In the case of V473\,Tau, since the magnetic structure of the star has already known, its position on the H-R diagram represents that it produces magnetic fields before reaching or in the beginning of the main sequence.

\section{Acknowledgments}
We acknowledge assistance from Vino Sangaralingam and Gemma Whittaker in the production of the data used in this study. The {\sl STEREO} Heliospheric imager was developed by a collaboration that included the Rutherford Appleton Laboratory and the University of Birmingham, both in the United Kingdom, and the Centre Spatial de Lige (CSL), Belgium, and the US Naval Research Laboratory (NRL),Washington DC, USA. The {\sl STEREO}/SECCHI project is an international consortium of the Naval Research Laboratory (USA), Lockheed Martin Solar and Astrophysics Lab (USA), NASA Goddard  Space Flight Center (USA), Rutherford Appleton Laboratory (UK), University of Birmingham (UK), Max-Planck-Institut fr Sonnen-systemforschung (Germany), Centre Spatial de Lige (Belgium), Institut dOptique Thorique et Applique (France) and Institut dAstrophysique Spatiale (France).

\references

Burke, E.W., Jr., Rolland, W.W., and Boy, W.R.: 1970, {\it Journal of the Royal Astronomical Society of Canada} {\bf 64}, 353  \BIBCODE{1970JRASC..64..353B}

Dubath, P., Rimoldini, L., S{\"u}veges, M., Blomme, J., L{\'o}pez, M., Sarro, L.M., De Ridder, J., Cuypers, J., Guy, L., Lecoeur, I., Nienartowicz, K., Jan, A., Beck, M., Mowlavi, N., De Cat, P., Lebzelter, T., and Eyer, L.: 2011, {\it Monthly Notices of the Royal Astronomical Society} {\bf 414}, 2602  \BIBCODE{2011MNRAS.414.2602D}

Eyles, C.J., Harrison, R.A., Davis, C.J., Waltham, N.R., Shaughnessy, B.M., Mapson-Menard, H.C.A., Bewsher, D., Crothers, S.R., Davies, J.A., Simnett, G.M., Howard, R.A., Moses, J.D., Newmark, J.S., Socker, D.G., Halain, J.-P., Defise, J.-M., Mazy, E., and Rochus, P.: 2009, {\it Solar Phys.} {\bf 254}, 387  \BIBCODE{2009SoPh..254..387E}

Ghosh, P.: 2007, {\it Rotation And Accretion Powered Pulsars. Series: World Scientific Series in Astronomy and Astrophysics, ISBN: 978-981-02-4744-7. WORLD SCIENTIFIC, Edited by Pranab Ghosh, vol. 10} {\bf 10} \BIBCODE{2007WSSAA..10.....G}

Hansen, C.J.~and Kawaler, S.D.: 1994, {\it Science} {\bf 265}, 1902. \BIBCODE{1994Sci...265.1902H}

Hubrig, S.~and Mathys, G.: 1994, {\it Astronomische Nachrichten} {\bf 315}, 343  \BIBCODE{1994AN....315..343H}

Hubrig, S., North, P., and Mathys, G.: 2000, {\it Astrophys. J.} {\bf 539}, 352   \BIBCODE{2000ApJ...539..352H}

Hubrig, S., Nesvacil, N., Sch{\"o}ller, M., North, P., Mathys, G., Kurtz, D.W., Wolff, B., Szeifert, T., Cunha, M.S., and Elkin, V.G.: 2005, {\it Astron. Astroph.} {\bf 440}, L37 \BIBCODE{2005A&A...440L..37H}

Jerzykiewicz, M.: 2009, {\it Acta Astronomica} {\bf 59}, 307  \BIBCODE{2009AcA....59..307J}

Kochukhov, O.: 2011, {\it Physics of Sun and Star Spots} {\bf 273}, 249   \BIBCODE{2011IAUS..273..249K}

Koupelis, T. and Kuhn, K. F.: 2007, {\it In Quest of the Universe. ISBN. 978-0763708108, Jones \& Bartlett Publishers, Sudbury}.

Maitzen, H.M.: 1977, {\it Astron. Astroph.} {\bf 60}, L29  \BIBCODE{1977A&A....60L..29M}

Mikulasek, Z., Szasz, G., Krticka, J., Zverko, J., Ziznovsky, J., Zejda, M., and Graf, T.: 2009, {\it ArXiv e-prints}, arXiv:0905.2565 \BIBCODE{2009arXiv0905.2565M}

Mikul{\'a}{\v s}ek, Z., Krti{\v c}ka, J., Henry, G.W., de Villiers, S.N., Paunzen, E., and Zejda, M.: 2010, {\it Astron. Astroph.} {\bf 511}, L7   \BIBCODE{2010A&A...511L...7M}

Mikul{\'a}{\v s}ek, Z., Krti{\v c}ka, J., Jan{\'{\i}}k, J., Zejda, M., Henry, G.W., Paunzen, E., {\v Z}i{\v z}{\v n}ovsk{\'y}, J., and Zverko, J.: 2014, {\it Putting A Stars into Context: Evolution, Environment, and Related Stars}, 270  \BIBCODE{2014psce.conf..270M} 

Oetken, L.: 1985, {\it Astronomische Nachrichten} {\bf 306}, 187   \BIBCODE{1985AN....306..187O}

Rakosch, K.D.~and Fiedler, W.: 1978, {\it Astronomy and Astrophysics Supplement Series} {\bf 31}, \BIBCODE{83 1978A&AS...31...83R}

Rimoldini, L., Dubath, P., S{\"u}veges, M., L{\'o}pez, M., Sarro, L.M., Blomme, J., De Ridder, J., Cuypers, J., Guy, L., Mowlavi, N., Lecoeur-Ta{\"i}bi, I., Beck, M., Jan, A., Nienartowicz, K., Ord{\'o}{\~n}ez-Blanco, D., Lebzelter, T., and Eyer, L.: 2012, {\it Monthly Notices of the Royal Astronomical Society} {\bf 427}, 2917   \BIBCODE{2012MNRAS.427.2917R}

Sangaralingam, V.~and Stevens, I.R.: 2011, {\it Monthly Notices of the Royal Astronomical Society} {\bf 418}, 1325   \BIBCODE{2011MNRAS.418.1325S}

Schaller, G., Schaerer, D., Meynet, G., and Maeder, A.: 1992, {\it Astronomy and Astrophysics Supplement Series} {\bf 96}, 269  \BIBCODE{1992A&AS...96..269S}

St{\c e}pie{\'n}, K.: 1998, {\it Astron. Astroph.} {\bf 337}, 754   \BIBCODE{1998A&A...337..754S}

Whittaker, G., Sangaralingam, V., and Stevens, I.: 2012, {\it From Interacting Binaries to Exoplanets: Essential Modeling Tools} {\bf 282}, 143 \BIBCODE{2012IAUS..282..143W}

Wraight, K.T., Fossati, L., Netopil, M., Paunzen, E., Rode-Paunzen, M., Bewsher, D., Norton, A.J., and White, G.J.: 2012, {\it Monthly Notices of the Royal Astronomical Society} {\bf 420}, 757  \BIBCODE{2012MNRAS.420..757W}

\endreferences

\end{document}